\documentclass[doublecol]{epl2} 
\usepackage[usenames,dvipsnames,table]{xcolor}
\usepackage[
  breaklinks,             %
  bookmarks = false, %
  pdfpagemode   = UseNone,%
  pdfstartview 	= FitH,		%
  pdfstartpage  = 1,      %
  colorlinks    = true   %
]{hyperref}
\hypersetup{
    linkcolor=RubineRed,          
    citecolor=ForestGreen,        
    filecolor=Mulberry,      
    urlcolor=RoyalBlue           
}
\usepackage{microtype}
\usepackage{graphicx}

\title{Correlation effects and spin-orbit interactions in two-dimensional hexagonal $5d$ transition metal carbides, $\mathrm{Ta}_{n+1}\mathrm{C}_n$ ($n=1,2,3$)}
\shorttitle{Electronic structure and magnetism in $5d$ transition metal carbides} 
\author{Nina J.\ Lane\inst{1} \and Michel W.\ Barsoum\inst{1} \and James M.\ Rondinelli\inst{1}\thanks{E-mail: \email{jrondinelli@coe.drexel.edu}}}
\institute{              
  \inst{1} Department of Materials Science \& Engineering, 
Drexel University, Philadelphia, PA 19104, USA
}%
\shortauthor{N.\ J.\ Lane \etal}

\abstract{
Density functional calculations are used to investigate the 
electronic structure of two-dimensional 5$d$ tantalum carbides with 
honeycomb-like lattice structures.  
We focus on changes in the low-energy bands near the Fermi level 
with dimensionality.
We find that the Ta 5$d$ states dominate, and
the extended nature of the wavefunctions makes them weakly correlated.  
The carbide sheets are prone to long range magnetic order and  
we evaluate their stability to enhanced electron--electron 
interactions through a Hubbard $U$ correction. 
Lastly, we find that the splitting of the bands near the Fermi level
caused by spin-orbit interaction decreases with increasing dimensionality. 
In the lowest dimensionality ($n=1$) case, the band splitting pushes a conduction 
band above the Fermi level and leads to a semi-metallic band
structure.
}

\begin{document}
\maketitle

\begin{figure}
\centering
\includegraphics[width=\columnwidth]{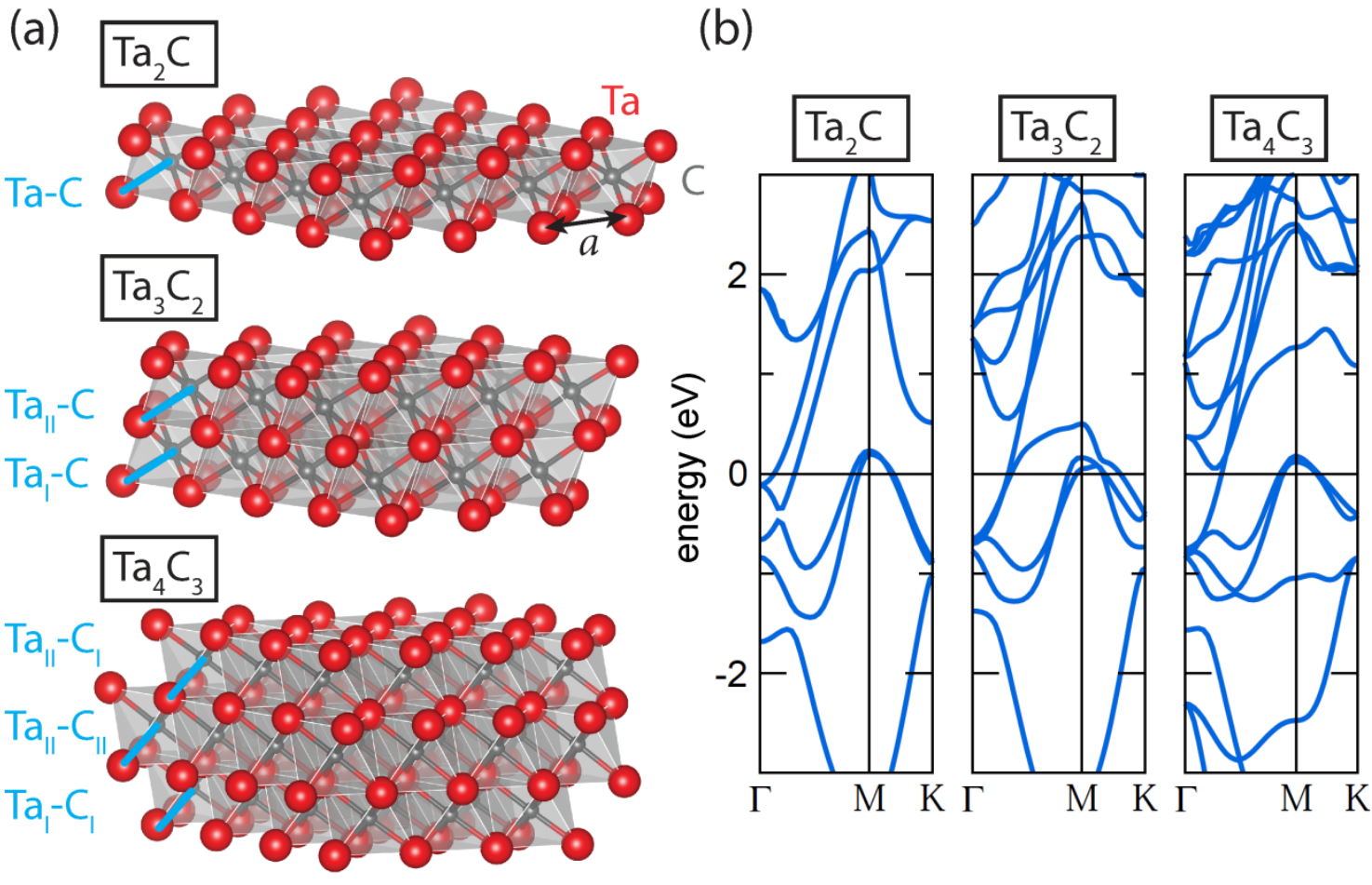}
\caption{The two-dimensional Ta$_{n+1}$C$_n$ ($n=1,2,3$) 
sheets (a) all possess hexagonal symmetry. 
The lattice constants, $a$, 
and select interatomic 
Ta--C distances are highlighted (cf.\ \autoref{table:structure}). 
(b) LDA electronic band structures for each compound 
along the path  $\Gamma (0,0,0) \rightarrow 
M (\frac{1}{2},0,0) \rightarrow K (\frac{1}{2},\frac{1}{2},0)$.
}
\label{fig:sb}
\end{figure}

 Two-dimensional (2D) free-standing crystals 
exhibit a range of functional properties, mainly derived from the 
topology of their underlying lattice and enhanced electronic 
and magnetic effects due to reduced dimensionality.
Spin-polarized edge states,\cite{qshgraphene} for example, 
have been predicted for the well-studied 2D carbon material 
graphene,\cite{graphene2004} owing to the topological origin of its 
transport properties. 
A large external magnetic field, however, is required to realize the 
quantum Hall effect, and its spin degeneracy 
makes it difficult to manipulate.  
To overcome these challenges, experimental approaches have been 
developed to induce magnetism by introducing transition metal adatoms 
\cite{KrasheninnikovTMembedding} and point defects \cite{Crvenkapointdefects} 
on the surfaces.
2D binary metal oxides and dichalcogenides, 
e.g.\  ZnO, BN, MoS$_2$ also find widespread interest. 
New functionalities originate from the presence of 
more diverse chemistries.
\cite{SeayadHstorage,MagrapheneMoSe2} 
However, in most existing pristine 2D free standing materials, 
magnetic ordering is absent and  the tunability of the electronic 
structure is limited to electrostatic doping.
A more promising avenue includes directly incorporating transition metals 
with multiple orbital degrees of freedom and highly-correlated electrons 
into to the lattice. 
Magnetism, for example, was recently predicted for 
V$X_2$ ($X$=S, Se) monolayers.\cite{MaVX2monolayers}
%
Alternatively, heavier $5d$ transition metals 
with strong spin-orbit coupling can be either deposited on 
the surface or directly integrated into the lattice of the 2D materials.
Recent first principles calculations show 
that graphene decorated with $5d$ transition 
metals can exhibit remarkable magnetic and topological transport properties
 \cite{PhysRevLettTunableTM}. The magnetic coupling induced by such treatments on nonmagnetic 2D materials, however,  
is difficult to control in actual applications due to unintentional impurities and defects.
In this Letter, we focus on low-dimensional hexagonal  materials,  
consisting of alternating layers of carbon and tantalum.  
These Ta-containing transition metal carbides are part of a recently discovered group of 
2D materials called ``MXenes'' synthesized by chemical 
exfoliation.\cite{MXeneorig}  
Similar to the previously studied Ta-decorated  graphene structures,
\cite{PhysRevLettTunableTM} these materials contain sheets of 
carbon in the inner layers and Ta atoms on the surface.  
In this case, however, the Ta layers are ordered, rendering them 
less susceptible to defects and more favorable for 
deliberate  surface functionalization. 
Furthermore,  these materials are derived from MAX phases,  which are a large family of layered carbides and nitrides with the general formula $M_{n+1}AX_n$, where $n=1\cdots3$,  
$M$ is an early transition metal, $A$ is an A-group element (mostly from groups 13 and 14),  and $X$ is carbon and/or nitrogen. \cite{MAX2000}
Since the stacking sequence of the $X\mathrm{M}_6$ octahedra  in the 
hexagonal MAX phases depends on the stoichiometry, the MXene sheets 
have the advantage that dimensionality controls both the system size and 
the symmetry between the two surfaces [Fig.~\ref{fig:sb}(a)]. 


The 2D MXenes have the general formula $M_{n+1}X_n$,  
and they crystallize in sheets containing 1, 2, or 3 layers of 
$X\mathrm{M}_6$ octahedra depending on $n$.  
Their recent synthesis \cite{MXeneorig} has 
spawned interest for uses in energy applications, including anodes in Li ion batteries \cite{MXeneLi,come} 
and electrodes for supercapacitors.  
There are a number of recent theoretical studies,
\cite{tanq,shein,kurt}
mainly focusing on the Ti-containing phases.   
However, modeling work on MXenes in the Ta-C system have 
not yet been reported, despite their recent synthesis \cite{MXenefamily}.
Motivated by the structural flexibility, possible enhanced 
electron--electron interactions, and the strong 
spin-orbit coupling parameter for 
Ta ($\zeta_d=1970$~cm$^{-1}$)\cite{pchandbook}, we  use 
\textit{ab initio} electronic 
structure calculations to investigate the 
effect of dimensionality ($n$) and electron correlations  
on the band structure and magnetic ordering in 
Ta$_{n+1}$C$_n$, $n=1\cdots3$.
First-principles density functional calculations are performed using the 
Vienna \textit{Ab initio} Simulation Package (\textsc{vasp}) \cite{vasp3}, 
with a plane wave cutoff of 500~eV and  the projector-augmented 
wave method (PAW)\cite{PAW}  to treat the interaction 
between the core and valence electrons; we 
treat the Ta 5$p$ electrons  as valence electrons.
For the site-decomposed density of states, partial occupations are set 
using the tetrahedron method with Bl\"{o}chl corrections. In the
band structure calculations, Gaussian smearing with a
smearing width of 0.10 eV was used.
Reciprocal space integrations are performed using a  $15\times15\times2$
$k$-point mesh.
We investigate the effects of electron--electron interactions by using  
both the local (spin) density approximation [L(S)DA)] and the improved 
generalized gradient approximation (GGA) of
Perdew-Burke-Eruzerhof (PBEsol) for solids\cite{PhysRevLett.100.136406} 
with the rotationally invariant Hubbard $U$ correction ($+U$) of Liechtenstein \textit{et al}.\cite{LDAUmethod}  \revision{For the on-site exchange interaction, we test values of $J$ from 0.2 to 1.5~eV and
determine $J$ does not have a strong effect on the stability of the magnetic configurations with $U$.  $J$ is therefore kept constant at 0.5~eV throughout.}  
The high measured conductivity of Ta$_4$C$_3$\cite{MXenefamily}, and the robustly metallic electronic structures of bulk TaC\cite{Li2TMC} and the MAX phases \cite{MAX2000}, suggest that electron correlation effects should be weak.  
We, therefore, anticipate the LDA and PBEsol functionals to provide 
an adequate description of the electronic and magnetic properties of these materials.
The Ta$_{n+1}$C$_n$ unit cells used in our calculations 
contain two symmetry equivalent free-standing sheets 
that are separated by 11-13~\AA~of vacuum. 
We obtain the equilibrium structures at the LDA and PBEsol level 
by minimization of the total energy 
computed for a range of $a$ lattice parameters, performing a 
full relaxation of the atomic positions along the $c$-direction until 
the forces are converged below a
tolerance of 5~meV~\AA$^{-1}$.

\begin{table}
\caption{\label{table:structure}Summary of lattice parameters, $a$,
and Ta--C bond lengths, $d$, in {\AA} 
with respect to dimensionality, $n$, obtained with 
LDA and PBEsol functionals. Embolden values correspond to 
experimental data taken from Ref.~\cite{MXenefamily}.}
\begin{tabular}{lllll}
\hline\hline
	& $n$	& length &	LDA			&PBEsol\\
\hline
Ta$_2$C& 1 & $a$ &3.041		&3.058\\
&&$d$(Ta--C)		&2.127		&2.139\\[0.1em]
\hline\\[-0.8em]
Ta$_3$C$_2$ &2 &$a$	&3.086		&3.112\\
&&$d$(Ta$_\mathrm{I}$--C) & 	2.110&		2.127\\
&&$d$(Ta$_\mathrm{II}$--C)& 	2.220&		2.236\\[0.1em]
\hline\\[-0.8em]
Ta$_4$C$_3$ &3&$a$&3.077		&3.094 $~~$(\textbf{3.1})\\
&&$d$(Ta$_\mathrm{I}$--C) &2.119		&2.131\\
&&$d$(Ta$_\mathrm{II}$--C$_\mathrm{I}$) &2.201	&2.210\\
&&$d$(Ta$_\mathrm{II}$--C$_\mathrm{II}$) &2.215	&2.226\\
\hline\hline
\end{tabular}
\end{table}

\autoref{table:structure} contains the ground state atomic structure 
descriptors obtained with the LDA and PBEsol functionals.
{These ground state atomic structures are used for all subsequent band structure 
and energy calculations according to the corresponding dimensionality and functional.}
The LDA functional, for all values of $n$, predicts equilibrium 
lattice parameters that are smaller than those obtained with PBEsol;
nonetheless, both functionals are in good agreement, within $<1\%$ 
of each other and the available experimental data\cite{MXenefamily} 
for Ta$_4$C$_3$.
We also summarize the interatomic distances between the different Ta and 
C atoms corresponding to the sites labeled in Fig.~\ref{fig:sb}(a).
For all 2D sheets explored, the Ta atoms at the 
surface layer have shorter Ta--C bonds than those in the center 
of the sheet, i.e. 
$d$(Ta$_\textrm{I}$--C) $< d$(Ta$_\textrm{II}$--C), which is also consistent with the 
bond lengths in MAX phases \cite{MAX2000}.
%

\begin{figure}
\centering
\includegraphics[width=1\columnwidth]{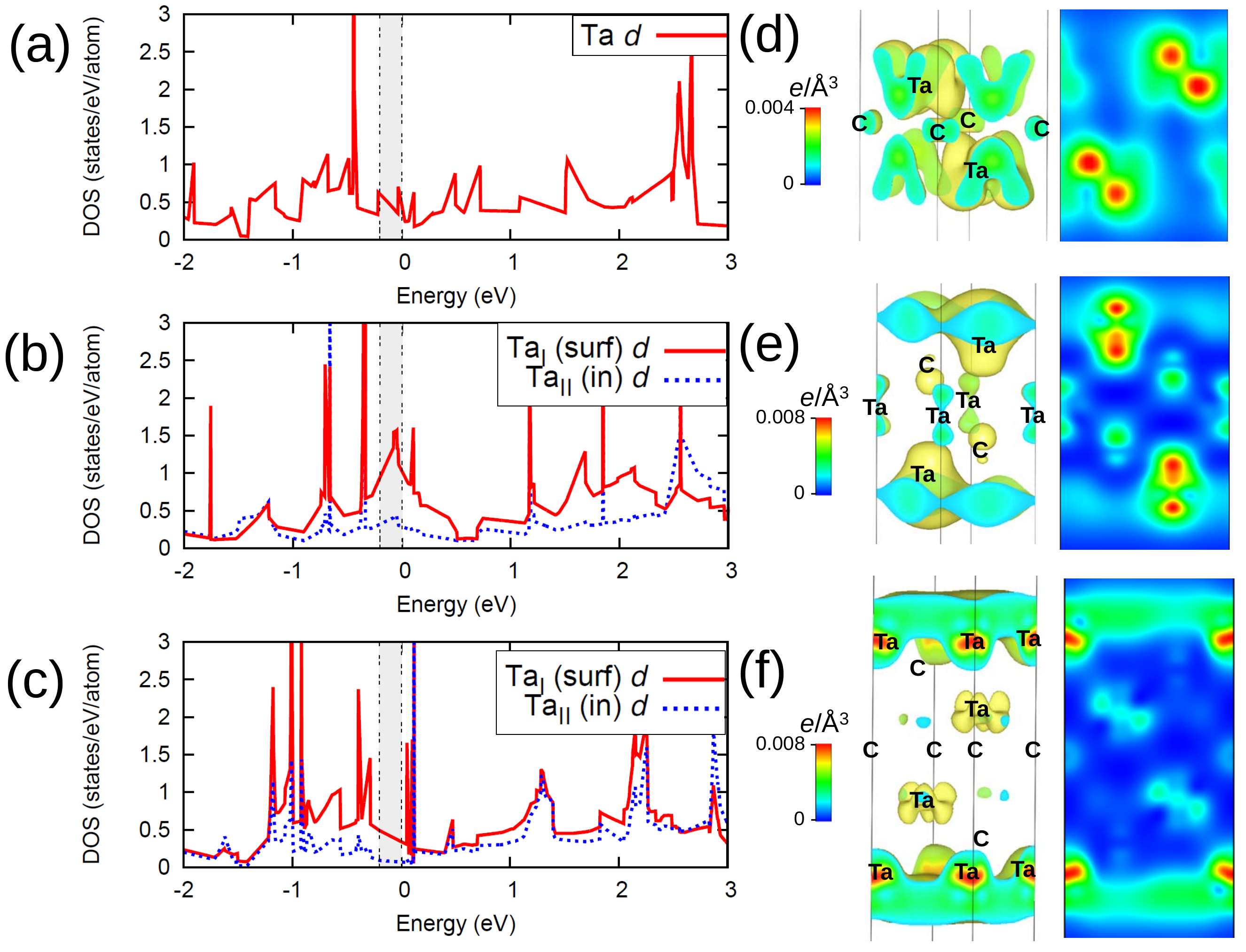}\vspace{3pt}
\caption{Site-decomposed partial DOS computed within the 
LDA for the surface (surf) and inner (in) layer Ta sites in (a) Ta$_2$C, (b) Ta$_3$C$_2$, 
and (c) Ta$_4$C$_3$. 
Note that the contribution from the C atoms to the DOS 
within this energy window is small and therefore not shown. 
The spatial distribution of the partial charge density from 0.2 eV up to 
the Fermi level (shaded region) is also shown in (d), (e), and (f) for Ta$_2$C, Ta$_3$C$_2$, 
and Ta$_4$C$_3$, respectively.
The 2D contours are projections on the (11\={2}0) plane.  \vspace{-8pt}}
\label{fig:db}
\end{figure}
While there are slight differences in the lattice dimensions and 
atomic positions, the average interatomic distances and lattice 
parameters are similar between the three stoichiometries. 
This suggests that any differences in electronic structure should  
originate from either the stacking of the octahedra or the number 
of occupied Ta $d$ bands available, 
which depends on the ratio of Ta to C atoms.   
To explore these possible differences, we begin by computing 
the electronic band structures with the LDA functional.  
For $n=1$, we find two dispersive Ta $d$-bands crossing the Fermi level ($E_F$) 
along $\Gamma - \textrm{M}$ [Fig. \ref{fig:sb}(b), left]. 
%
The two free-electron pockets centered at $\Gamma$ are 
similar to those found in 3D metals despite the 2D nature of the MXene
sheet.
{Further, the hole pockets at M make this a semi-metal with possible $p$-type conductor features that may be highly temperature dependent.
%

In contrast, in Ta$_3$C$_2$ ($n=2$) we find multiple bands crossings
at the Fermi level, with metallic-like partial occupancy 
centered around M}. 
%
%
%
%
{The band structure for Ta$_4$C$_3$ ($n=3$) shows similar
features as Ta$_2$C, with nearly fully occupied Ta $d$ states at M.  However,  
in this case a single Ta band, with nearly linear dispersion, crosses  
the Fermi level along $\Gamma - \textrm{M}$.} 
Indeed the site-decomposed partial densities-of-states (DOS) confirm
that the region near $E_F$ is largely controlled by the Ta $d$ states
[Figs.~\ref{fig:db}(a--c)].
Figs.~\ref{fig:db}(d--f) show the spatial distribution of the electrons within 
0.2 eV of $E_F$.
We find  a strong dependence on dimensionality for the charge distribution.
Intriguingly, this spatial  distribution about the Ta site 
in $n=1$, \textit{viz} Ta$_2$C  [Fig.~\ref{fig:db}(d)], and  the inner Ta atom in $n=3$, \textit{viz} Ta$_4$C$_3$  [Fig.~\ref{fig:db}(f)], 
share similar features---the charge around the atom is distributed into six lobes, three above and three below the Ta atom.  
The inner Ta atom in  $n=2$  (Ta$_3$C$_2$)  [Fig.~\ref{fig:db}(d)]  
shows strikingly different behavior, with a small distribution of 
charge collected above and below the Ta atom, aligned parallel to the 
$c$-axis.

\begin{figure}
\centering
\includegraphics[width=0.72\columnwidth]{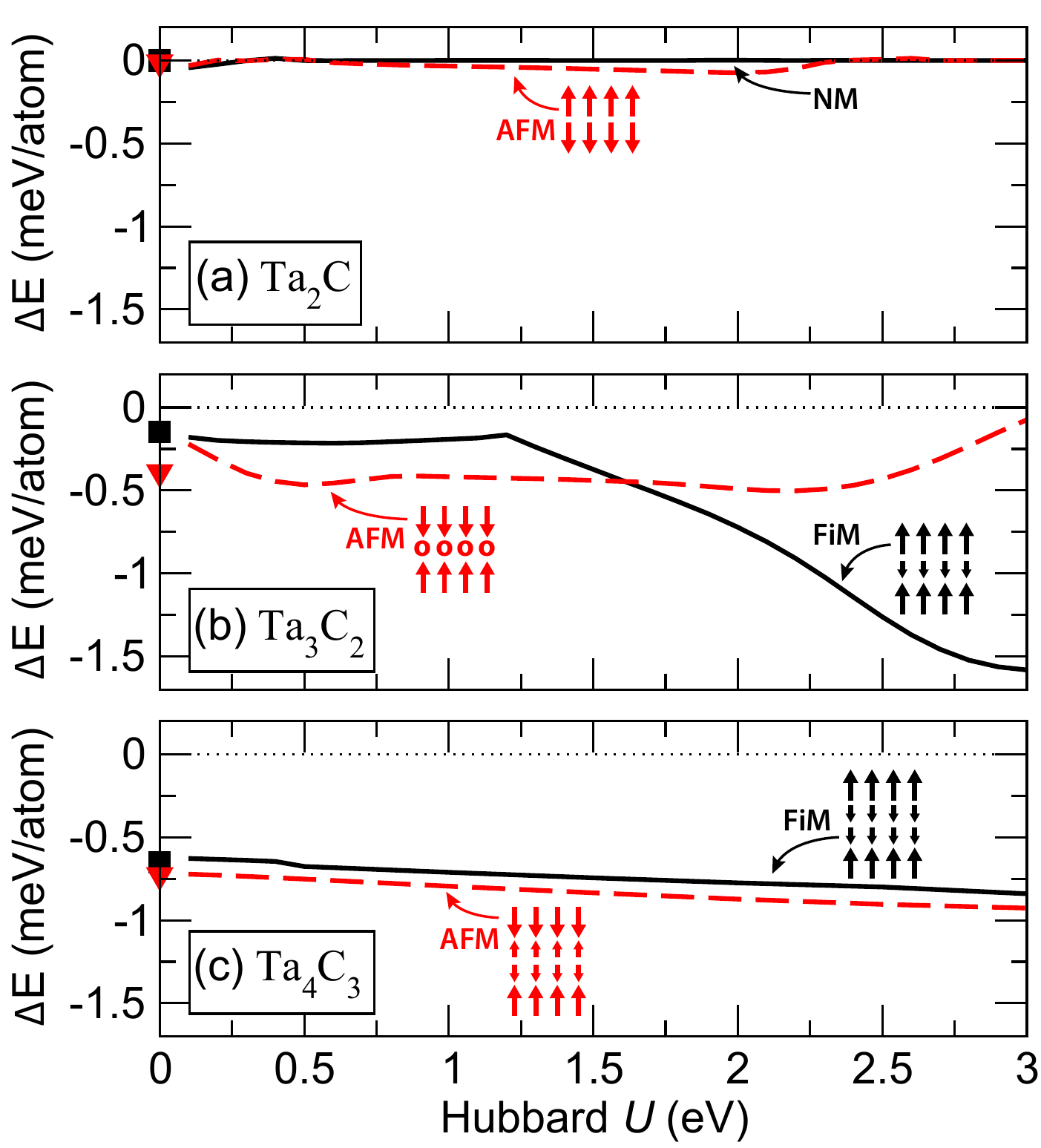}\vspace{-8pt}
\caption{
The L(S)DA$+U$ energy differences, $\Delta E$,  
between the magnetic and non-magnetic (NM)
states with $U$ 
for (a) Ta$_2$C, (b) Ta$_3$C$_2$, and (c) Ta$_4$C$_2$. 
Energies given per Ta atom.
The points at $U=0$~eV are obtained without the Hubbard $U$ method.
Schematics (inset) illustrate the spin ordering of the Ta atoms with 
ferrimagnetic (FiM) or  
antiferromagnetic (AFM) order, where  
the arrows represent the relative magnitudes and directions of spin, 
and the open circles indicate Ta atoms with no magnetic moment.  
\vspace{-8pt}}
\label{fig:sp}
\end{figure}

To understand the atomic-scale origin of these features, 
we examine more closely the crystal structures of each sheet.
Ta$_2$C ($n=1$) consists of a single Ta--C octahedron with stacking sequence AcB, where uppercase letters denote the Ta atom and lower case letters represent the stacking of carbon atoms.  Ta$_4$C$_3$ ($n=3$), 
therefore, is obtained as a Ta$_2$C layer, AcB, with an extra Ta atom on each surface -- that is, Cb[AcB]aC.  Note that this is the only structure in which two Ta atoms on either surface have the same stacking. On the other hand, Ta$_3$C$_2$ ($n=2$) has a stacking sequence of AcBaC, so the symmetry of the AcB layer is broken due to the odd number of layers. 
This suggests that the differences in the charge distribution we find 
in Figs.~\ref{fig:db}(d--f) are largely governed by the stacking sequence of the CTa$_6$ units---a degree of freedom unique to the MXene 
phases.
Such differences  in the electron distribution near $E_F$ due to the 
asymmetric stacking are also visible in the shape of the DOS [shaded region in Fig. \ref{fig:db}(b)].
The charge is concentrated at the surface for Ta$_3$C$_2$ and Ta$_4$C$_3$, consistent with the higher partial density of states for the Ta surface atoms [Figs.~\ref{fig:db}(b--c)].

The large number of Ta $d$-states at $E_F$ and the sensitivity of the 
electronic structure to the sheet dimensionality suggest 
the possibility of stable long range magnetic spin configurations. 
We therefore performed a series of spin-polarized calculations 
with two different starting configurations, corresponding to 
ferromagnetic (FM), ferrimagnetic (FiM), and antiferromagnetic (AFM) spin order on the 
Ta sites to systematically explore the possible magnetic orders with 
respect to $n$. 
We carried out unconstrained-spin density calculations to find 
the stable magnetic ordering (Fig. \ref{fig:sp})  
and compared the total energy of those states to that of the 
non-spin-polarized case.  
Given the limited ability of DFT to fully capture correlation effects, 
including transition metal ions with partially filled $d$ shells, we 
now add the Hubbard $U$ correction 
to the standard 
PBEsol (PBEsol$+U$) and LSDA (LSDA$+U$) functionals \cite{LDAU}.  
Since it was recently suggested that Ta-containing MAX phases are 
weakly correlated  \cite{correlatemax}, we explored $U < 3.0$ eV. 

\autoref{fig:sp} shows the change in total energy {of the spin-polarized states} computed with the LSDA$+U$ functional compared to the non-magnetic (NM) case.  
{
The change in energy is calculated as a function of $U$ by
$\Delta E(U) = [E_\mathrm{M}(U) - E_\mathrm{NM}(U)]/N$, 
where $E_\mathrm{M}(U)$ and  $E_\mathrm{NM}(U)$
are the total energies of the magnetic and non-magnetic states, 
respectively,
for a given $U$ value. $N$ is the number of atoms per unit cell.
}
In {most}  cases, the magnetically ordered configurations 
are lower in energy than the NM state. {For $n=1$, the FM state could not
be stabilized with LDA calculations, and weak magnetic
ordering is observed in the metastable AFM state [Fig. \ref{fig:sp}(a)].} As the dimensionality increases, the AFM case becomes more stable{.}
%
Our main results are summarized in Table \ref{table:spin}, where the values in parentheses specify the $U$ values above which the specific magnetic ordering becomes stable.
For Ta$_3$C$_2$, both LSDA$+U$ and PBEsol$+U$ 
calculations predict a FiM configuration, whereas an 
AFM ordering is predicted for Ta$_4$C$_3$, 
but only  with the LSDA$+U$ exchange-correlation functional (Table \ref{table:spin}). 
Note that in Ta$_4$C$_3$ no ordered magnetic phase was found 
to be stable with either PBEsol and PBEsol$+U$. 
 
\begin{table}[t]
\caption{\label{table:spin}Summary of the stable spin polarized 
ground states for the Ta$_{n+1}$C$_n$ MXene phases using different exchange-correlation functionals with and without a Hubbard $U$ correction. Notations (schematics) for the magnetic states are given in the caption (insets) 
to Fig.~\ref{fig:sp}. 
}
\begin{tabular}{llll} 
\hline\hline
& Ta$_2$C & Ta$_3$C$_2$ & Ta$_4$C$_3$ \\
\hline
LDA & NM & AFM & AFM \\
LDA+U & AFM ($U > 0.5$) & FiM ($U > 1.6$) & AFM \\
\hline\\[-0.8em]
PBEsol& AFM & AFM & NM \\
PBEsol+U & FM ($U>0.1$) & FiM ($U > 0.9$) & NM \\
\hline\hline
\end{tabular}
\end{table}

%
In all cases, the surface Ta atoms are spin polarized
in the same direction for FiM order and in opposite directions 
for AFM.   
The Ta atoms in the inner layers are weakly spin-polarized in the 
direction opposite to the surface Ta atoms in both the FiM and AFM configurations, with the exception of the AFM configuration of 
Ta$_3$C$_2$, which is constrained by symmetry. 
The addition of $U$ generally has a small effect on the electronic structure, 
leading to a slight shift of the Ta $d$-bands uniformly to higher energies.
%

\begin{figure}
\centering
\includegraphics[width=0.95\columnwidth]{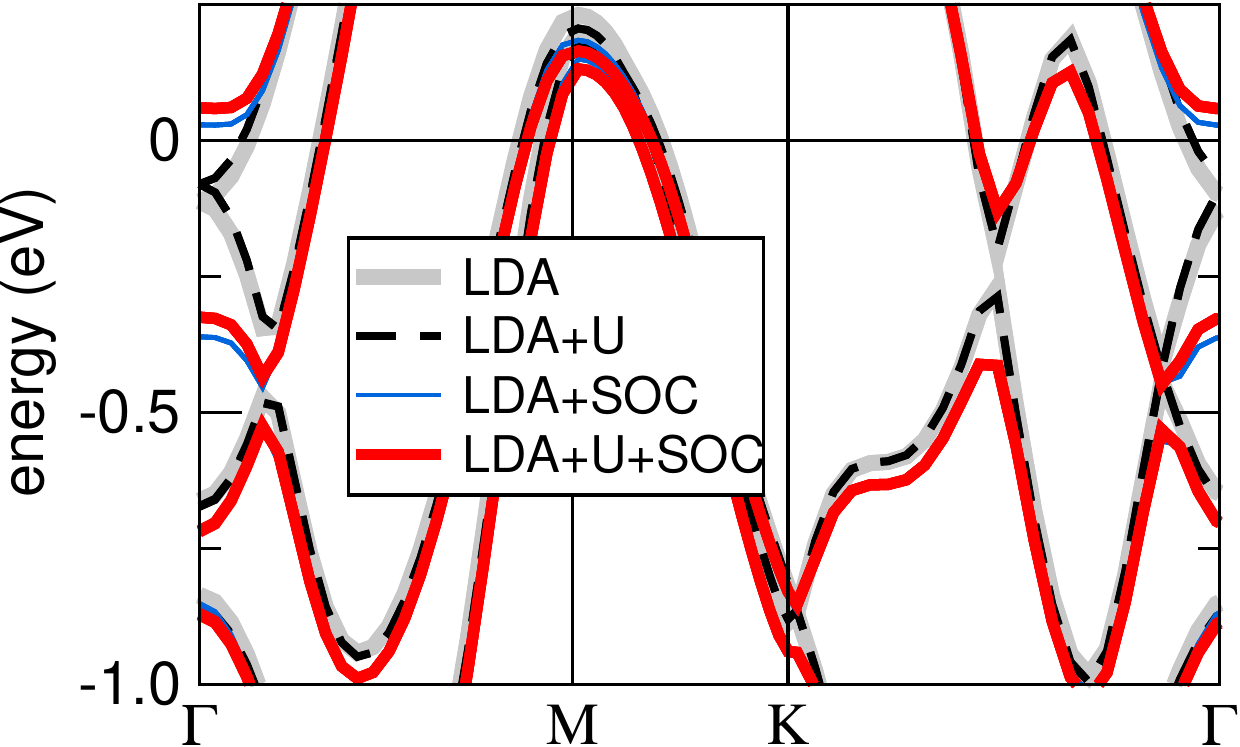}\vspace{-8pt}
\caption{LDA band structure for {Ta$_2$C without spin or correlation (gray, thick lines), using LDA+U with $U$=1eV (black, broken lines), with spin-orbit coupling (SOC) (blue, solid lines), and with SOC using LDA+U with $U$=1 eV (red, thick lines)}.
}
\label{fig:soc}
\end{figure}

We now evaluate the effect of spin-orbit coupling (SOC) on the 
band structure with respect to $n$. 
Here we find that SOC {splits the bands near the Fermi level, manifesting as a shift in energy, especially for the highest occupied Ta $d$-bands.  It has the strongest effect on Ta$_2$C (Fig. \ref{fig:soc}), where the band splitting is prominent and pushes one of the formerly occupied degenerate bands below $\approx-0.11$~eV to above $E_F$, driving a transition from a band structure with two band crossings to one with Fermi surface with small electron and hole pockets.}
{Including electron correlation (+U) within LDA and LDA+SOC causes only a small energy shift to those states (Fig. \ref{fig:soc}). 
For Ta$_2$C, the spin-orbit splitting at the top of the valence band at $\Gamma$ ($\Delta_{SO}$=269 meV) is comparable to} that observed in GaAs ($\Delta_{SO}$=342 meV)\cite{socgaas}, {and more than} 3 orders of magnitude larger than 
graphene ($\Delta_{SO}=\sim 0.05$ meV)\cite{PhysRevB.75.121402}.
{The result is that this double band crossing near $\Gamma$ shifts to a single linear band crossing as one band splits off and is pushed above the Fermi level.  This shift toward a more semi-metallic electronic state should be experimentally 
observed in its transport properties, which are predicted to be fundamentally
different from the $n=2$ and $n=3$ MXenes.} 

In summary, we have shown that the Ta-based 5$d$ electronic structure is sensitive to dimensionality.
All explored phases exhibit correlation stabilized magnetic order that is
not found in the bulk MAX phase structures.
{The LSDA$+U$ method stabilizes the ferromagnetic ordering in the case of $n=2$, 
and for $n=1$ spin-orbit coupling shifts the electronic band structure with a transition
from a two-band to a nearly filled single band.} 
In these 2D MXenes, the electronic structure is controlled by the 
stacking of the CTa$_6$ octahedra and the states derived from 
the surface Ta atoms. 
Tailoring the electronic structure could therefore be achieved 
through end group functionalization of the surfaces of the 
MXene sheets. 
This  opens up possibilities for engineering a class of tunable functional 2D materials.  
We conjecture that one could maintain 
a band structure {with linear graphene-like crossings 
where the Fermi surface is nearly completely 
controlled by a single band -- like that observed in {Ta$_2$C -- }}through epitaxial strain engineering.
Overall, the Ta-containing graphene-like carbides
show great promise as functional 2D materials
that can be synthesized in different dimensionalities,
leading to a range of stacking sequences and stoichiometries that
offer a variety of electronic and magnetic behaviors.

\acknowledgments
N.J.L.\ and M.W.B.\ were supported by the Assistant Secretary for Energy Efficiency and Renewable Energy, Office of Vehicle Technologies of the U.S.\ DOE under Contract No.\ DE-AC02-05CH11231, Subcontract 6951370 under the Batteries for Advanced Transportation Technologies (BATT) Program.  N.J.L. acknowledges financial support by the Integrated Graduate Education and Research Traineeship (IGERT) under NSF (DGE-0654313), and  
J.M.R.\ was supported by ARO (W911NF-12-1-0133).
This work benefited from the XSEDE (NSF) and the Center for Nanoscale Materials (U.S.\ DOE-BES, under Contract 
No.\ DE-AC02-06CH11357) HPC resources.

\bibliographystyle{eplbib}
\bibliography{refs}	
\end{document}